\documentclass[aps,prl,twocolumn,superscriptaddress,amsmath,amssymb,floatfix]{revtex4}  
\usepackage{graphicx}  
\usepackage{dcolumn}   
\usepackage{bm,color}        
\usepackage{amssymb}   

\newcommand{\be}{\begin{equation}}
\newcommand{\ee}{\end{equation}}

\newcommand{\R}{R}

\newcommand{\Ktwo}{{\cal K}_2}
\newcommand{\Kthree}{{\cal K}_3}
\newcommand{\s}{\sigma}
\newcommand{\gammah}{h}
\newcommand{\Atwo}{{A_2}}
\newcommand{\Athree}{{A_3}}
\newcommand{\Afour}{{B_4}}
\newcommand{\Afive}{{B_5}}
\newcommand{\Bfour}{{A_4}}
\newcommand{\Bfive}{{A_5}}
\newcommand{\Y}{{Y}}

\newcommand{\Gtwo}{G_2{}}
\newcommand{\Gthree}{G_3{}}
\newcommand{\Gfour}{G_4{}}
\newcommand{\Gfive}{G_5{}}

\newcommand{\Hfour}{F_4{}}
\newcommand{\Hfive}{F_5{}}

\allowdisplaybreaks

\begin{document}
 
\title{Healthy theories beyond Horndeski}
\author{J\'er\^ome Gleyzes}
\affiliation{CEA, Institut de Physique Th{\'e}orique,
         F-91191 Gif-sur-Yvette c{\'e}dex,  
         CNRS, Unit{\'e} de recherche associ{\'e}e-2306, F-91191 Gif-sur-Yvette c{\'e}dex}
\affiliation{Universit\'e Paris Sud, 15 rue George Cl\'emenceau, 91405,  Orsay, France}
\author{David Langlois} 
\affiliation{APC, (CNRS-Universit\'e Paris 7), 10 rue Alice Domon et L\'eonie Duquet, 75205 Paris, France,  
         }
\author{Federico Piazza}
\affiliation{APC, (CNRS-Universit\'e Paris 7), 10 rue Alice Domon et L\'eonie Duquet, 75205 Paris, France,  
         }
\affiliation{PCCP, 10 rue Alice Domon et L\'eonie Duquet, 75205 Paris, France}
\affiliation{CPT, Aix Marseille Universit\'e, UMR 7332, 13288 Marseille,  France.
}
\author{Filippo Vernizzi}
\affiliation{CEA, Institut de Physique Th{\'e}orique,
         F-91191 Gif-sur-Yvette c{\'e}dex,  
         CNRS, Unit{\'e} de recherche associ{\'e}e-2306, F-91191 Gif-sur-Yvette c{\'e}dex}
        \date{\today}

\begin{abstract}
We introduce a new class of scalar-tensor theories  of gravity that extend Horndeski, or ``generalized galileon", models. Despite possessing equations of motion of higher order in derivatives, we show that the true propagating degrees of freedom obey well-behaved second-order equations and are thus free from Ostrogradski instabilities,  in contrast to standard lore. Remarkably,  the covariant versions of the original galileon Lagrangians---obtained by direct replacement of derivatives with covariant derivatives---belong to this class of theories. These extensions of Horndeski theories exhibit an uncommon, interesting phenomenology: The scalar degree of freedom affects the speed of sound of matter, even when the latter
is minimally coupled to gravity. 
\end{abstract}

\maketitle

The discovery of the present cosmological acceleration has spurred the exploration of gravitational theories that could account for this effect. Many extensions of general relativity (GR) are based on the inclusion of a scalar degree of freedom (DOF)  in addition to the two tensor propagating modes of GR (see e.g.~\cite{Clifton:2011jh} for a review). In this context, a recent important proposal is the so-called galileon models~\cite{NRT}, with Lagrangians that involve second-order derivatives of the scalar field and lead, nevertheless, to equations of motions of second order. Such a property guarantees the avoidance of Ostrogradski instabilities, \emph{i.e.}~of the ghost-like DOF that are usually associated with higher time derivatives (see e.g.~\cite{woodard}).

Initially introduced in Minkowski spacetime, galileons have then been generalized to curved spacetimes \cite{Deffayet:2009wt,Deffayet:2009mn,Deffayet:2011gz}, where they turn out to be equivalent to a class of theories originally constructed by Horndeski forty years ago \cite{horndeski}.
Today, Horndeski theories, which  include quintessence, $k$-essence and $f(R)$ models,  constitute the main theoretical framework for scalar-tensor theories, in which cosmological observations are  interpreted.   
The purpose of this Letter is to show that this framework is not as exhaustive as generally believed, and can in fact be extended to include new Lagrangians. 
Indeed, having equations of motion of second order in derivatives---while indeed sufficient---is not necessary to avoid Ostrogradski instabilities, as already pointed out in e.g. \cite{deRham:2011qq,Zumalacarregui:2013pma}.
The theories beyond Horndeski that we propose lead to distinct observational effects and are thus fully relevant for an extensive comparison of scalar-tensor theories with observations.

\vskip.1cm
\emph{The model.}
The theories that we consider here can be viewed as a broader generalization of the galileons to curved spacetimes . They are described by  linear combinations of the Lagrangians 
\begin{align}
L_2^{\phi} & \equiv  \Gtwo(\phi,X)\;,  \label{L2} \\
L_3^{\phi} & \equiv  \Gthree(\phi, X) \, \Box \phi \;, \label{L3} \\
L_4^{\phi} & \equiv \Gfour(\phi,X) \, {}^{(4)}\!R - 2 \Gfour_{,X}(\phi,X) (\Box \phi^2 - \phi^{ \mu \nu} \phi_{ \mu \nu}) \nonumber \\
&+\Hfour(\phi,X) \epsilon^{\mu\nu\rho}_{\ \ \ \ \sigma}\, \epsilon^{\mu'\nu'\rho'\sigma}\phi_{\mu}\phi_{\mu'}\phi_{\nu\nu'}\phi_{\rho\rho'}\;, \label{L4} \\
L_5^{\phi} & \equiv \Gfive(\phi,X) \, {}^{(4)}\!G_{\mu \nu} \phi^{\mu \nu}  \nonumber \\
&+  \frac13  \Gfive_{,X} (\phi,X) (\Box \phi^3 - 3 \, \Box \phi \, \phi_{\mu \nu}\phi^{\mu \nu} + 2 \, \phi_{\mu \nu}  \phi^{\mu \sigma} \phi^{\nu}_{\  \sigma}) \nonumber \\ \;&
+\Hfive (\phi,X) \epsilon^{\mu\nu\rho\sigma}\epsilon^{\mu'\nu'\rho'\sigma'}\phi_{\mu}\phi_{\mu'}\phi_{\nu \nu'}\phi_{\rho\rho'}\phi_{\sigma\sigma'} \label{L5}\,,
\end{align}
which depend on a scalar field $\phi$ (and its derivatives $\phi_\mu\equiv \nabla_\mu \phi, \,  \phi_{\mu\nu}\equiv \nabla_{\nu}\nabla_\mu \phi$), on $X\equiv g^{\mu \nu}  \phi_\mu  \phi_\nu$, and on a metric $g_{\mu\nu}$ with respect to which matter is assumed to be minimally coupled;   $\epsilon_{\mu \nu \rho \sigma }$ is the totally antisymmetric Levi-Civita tensor and a comma denotes a partial derivative with respect to the argument.
Horndeski theories correspond to a subset of the above theories, subjected to the restricting conditions
\be
\label{horndeski_cond}
\Hfour(\phi,X) = 0 \,, \qquad \Hfive (\phi,X) =0 \,,
\ee
which ensure that the equations of motion (EOM) are second order. By contrast, we allow here arbitrary functions 
$\Hfour$ and $\Hfive$, which means that our theories contain two additional free functions with respect to the Horndeski ones. 

The new terms proportional to $\Hfour$ and  $\Hfive $ are, respectively, the covariant version of the original quartic and quintic galileon Lagrangians proposed in Ref.~\cite{NRT}. This guarantees second-order dynamics for the scalar field in the absence of gravity. When the metric is dynamical, the EOM involve up to third-order derivatives in these extended theories, but this does not imply the presence of unwanted extra DOF,  as we show below.

\vskip.1cm
\emph{Arnowitt-Deser-Misner (ADM) formulation.} 
In cosmology, where the scalar field gradient is timelike, it is convenient to perform an ADM decomposition of spacetime,  with metric
\be
ds^2=- N^2 dt^2  +\gammah_{ij} (dx^i+ N^i dt)(dx^j+ N^j dt ) \;,
\ee
by choosing the uniform scalar field ($\phi=$ const)  hypersurfaces
as constant-time hypersurfaces.
The above Lagrangians then have a very simple form in terms of the  intrinsic and extrinsic 3-d curvature tensors of the spatial slices, $R_{ij}$ and $K_{ij}$, as well as the lapse function $N$. This reformulation uses the unit  vector  $n^\mu \equiv - \phi^\mu/\sqrt{-X}$ normal to the uniform $\phi$  hypersurfaces, in terms of which the extrinsic curvature  is given by $K_{\mu \nu} \equiv (g^\sigma_{\ \mu} + n^\sigma n_\mu ) \nabla_\sigma n_\nu$. We also make use of the Gauss-Codazzi relations to relate the 4-d curvature to the 3-d one.

After cumbersome but straightforward manipulations, one finds that any combination of the $L_a^{\phi}$ leads to an ADM  Lagrangian density of the form ${\cal L} = \sqrt{-g} \sum_a L_a $, with
\begin{align}
&L_2  =\Atwo\;, \qquad L_3=\Athree\;   K \;, \nonumber \\
&L_4 = \Bfour \;  \Ktwo + \Afour \;   \R\;, \label{action} \\
&L_5 = \Bfive \; \Kthree  + \Afive\;   K^{ij} \;  \big[\R_{ij} -\gammah_{ij} R/2 \big]  \nonumber \;,
\end{align}
where 
$K\equiv h^{ij}K_{ij}$, $R\equiv h^{ij}R_{ij}$, and the quantities ${\cal K}_2$ and ${\cal K}_3$ are, respectively, quadratic and cubic combinations of 
$K_{ij}\equiv(\dot h_{ij}-D_iN_j-D_jN_i)/(2N)$ (where $D_i$ is the covariant derivative of $h_{ij}$), explicitly defined as
\begin{align}
\Ktwo & \equiv K^2 - K_{ij}K^{ij} \;, \label{K2} \\
\Kthree & \equiv K^3 - 3 K K_{i j}K^{i j} + 2  K_{i j}  K^{i k} K^j_{\ k} \;. \label{K3}
\end{align}
The coefficients in eq.~\eqref{action} are related to the original functions in eqs.~\eqref{L2}--\eqref{L5} by
\begin{align}
A_2 &= G_2 - \textstyle{(-X)^\frac12 \int \frac{G_{3,\phi}}{2\sqrt{-X}} dX}\, ,\\
A_3 &= - \textstyle{\int G_{3,X}\sqrt{-X} dX - 2 \sqrt{-X}G_{4,\phi}}\, ,\\
A_4 & = -G_4 + 2 X G_{4,X}+ \textstyle{\frac{X}{2}G_{5,\phi} - X^2 F_4}\, ,\label{A4}\\
B_4 & = G_4 + \textstyle{\sqrt{-X}\int \frac{G_{5,\phi}}{4 \sqrt{-X}} dX} \, ,\\
A_5 & = \textstyle{-\frac{(-X)^\frac32}{3} G_{5,X} + (-X)^\frac52 F_5}\, ,\\
B_5 & = - \textstyle{\int G_{5,X} \sqrt{-X} dX}\label{B5}\, .
\end{align}
In this ADM formulation, these  functions of $\phi$ and $X$ can also be seen as functions of $t$ and $N$ via the relations
 $\phi = \phi_0(t)$  and $X= -\dot\phi_0^2(t)/N^2$.

By using eqs.~\eqref{A4}--\eqref{B5}, the Horndeski conditions (\ref{horndeski_cond}) translate into 
\be
\Bfour= - \Afour +2 X \Afour_{,X}\;, \qquad \Bfive =  - X \Afive_{,X}/3\; \label{gal}.
\ee

\vskip.1cm
\emph{Hamiltonian analysis.}
In general, higher derivative theories are pathological because they lead, according to Ostrogradski's theorem, to \emph{extra} DOF that behave like ghosts.
Here we show, by resorting to a simple counting of the number of DOF in the Hamiltonian formalism, that the theories (\ref{action}) do not contain more than three degrees of freedom. Thus, there is no room for an extra DOF in addition to the scalar DOF initially built in and the two tensor modes similar to those of GR.

The Hamiltonian is obtained from the Lagrangian via a Legendre transform,
\be
H=\int d^3 x \left[\pi^{ij} \dot\gammah_{ij} -{\cal L}\right],
\ee
where the $\pi^{ij}$ are the conjugate momenta associated with the $\gammah_{ij}$, defined by
\be \label{invert}
\pi^{ij}=\frac{\partial {\cal L}}{\partial \dot\gammah_{ij}}\,.
\ee
Ignoring $L_5$ for simplicity, one can easily invert the above relation to express $\dot\gammah_{ij}$ as a function of $\pi^{ij}$ and obtain the explicit Hamiltonian, which can be written in the form
\be
H=\int d^3x \left[ N {\cal H}_0+N^i{\cal H}_i\right]\,,
\ee
with
\begin{align}
{\cal H}_0\equiv &-\sqrt{h} \Big[\Atwo - \frac{3 A_3^2}{8 A_4}  +\frac{\Athree \pi}{2 \sqrt{h} \Bfour}  + \Afour \R\nonumber \\
&+ \frac1{2 h \Bfour } \big(2\pi_{ij} \pi^{ij}-\pi^2 \big)\Big] \;, \\
{\cal H}_i\equiv& -2 D_j \pi_{\ i}^j \, .
\end{align}
We leave aside the uninteresting case $A_4=0$, which does not contain propagating tensor DOF.

In GR, variation with respect to $N$ and $N^i$ yields, respectively, the Hamiltonian constraint ${\cal H}_0=0$ and the momentum constraints ${\cal H}_i=0$. These constraints are, in Dirac's terminology, first class and eventually eliminate eight  out of the initial ten degrees of freedom (see e.g.~\cite{Henneaux:1992ig}). In our case, the gauge invariance under {\it spatial} diffeomorphims is preserved, leading to  first-class constraints analogous to the momentum constraints of GR and eliminating six DOF (see \cite{GLPV3} for details). However, variation with respect to $N$ now gives the constraint $\tilde{\cal H}_0\equiv{\cal H}_0+N\partial {\cal H}_0/\partial N=0$, which is in general second class, instead of first class. This can be understood as a consequence of  the scalar field that fixes the preferred slicing and thus breaks the full spacetime diffeomorphism invariance. This entails the elimination of only one DOF (instead of two in GR). 
Note that  this reasoning crucially depends on the absence of  $\dot N$ from the Lagrangians \eqref{action}, which is guaranteed by the specific form of the new terms proportional to $F_4$ and $F_5$ introduced in eqs.~\eqref{L4} and \eqref{L5}.
The final number of physical DOF is therefore three, which correspond to the two standard tensor modes plus a scalar mode, as will be clear from the linear analysis below.

When $L_5$ is included,  the full  Hamiltonian cannot be written in closed form because one cannot invert explicitly the relation~\eqref{invert}, even if the inversion is in general well defined locally \cite{GLPV3}. For this reason, we have not been able to  compute explicitly  the constraint algebra in the full case. However, our counting depends only on the nature of the constraints. Since the full Hamiltonian  is, by construction, invariant under spatial diffeomorphims, the associated constraints should remain first class and thus  eliminate six DOF as before. Taking into account the other constraints, one thus expects at most three DOF and, therefore, the absence of any ghostly extra DOF.  The counting is also similar if one includes matter, with the matter DOF adding to the three from the gravitational sector.
Finally,  note that our analysis could also be applied almost straightforwardly  to general ADM Lagrangians invariant under spatial diffeomorphisms involving arbitrary combinations of the extrinsic and intrinsic curvature tensors and their spatial derivatives. However, such a wider set of possibilities is not necessarily a covariant extension of galileons as eqs.~\eqref{L2}--\eqref{L5}. 

\vskip.1cm
\emph{Covariant formulation.} 
The above Hamiltonian analysis is based on our ADM reformulation of the theories and requires the  gradient of the scalar field to be timelike so that uniform scalar-field hypersurfaces are spacelike. Although this is the case in cosmology, which is the main motivation to study these models, one can wonder whether our findings are still valid for more general situations. 

For simplicity, let us consider theories involving up to  $L_4^\phi$, but not $L_5^\phi$. We have found that the  analysis of their equations of motion can be greatly simplified via the use of disformal transformations. Indeed,  the gravitational action with the Lagrangians \eqref{L2}--\eqref{L4} reexpressed in terms of $\phi$ and of the new metric 
\be
\label{transg}
\tilde g_{\mu\nu}=g_{\mu\nu}+ \Gamma(\phi, X) \partial_\mu\phi \partial_\nu\phi\, ,
\ee
with   
\be
\Gamma=\int\frac{F_4}{G_4-2XG_{4X}+X^2F_4} dX\, ,
\ee
turns out to  belong to the Horndeski class. This means that the equations of motion obtained by varying the action with respect to the metric $\tilde g_{\mu\nu}$ are second order. By using this property and by combining the (third-order) equations of motion for $\phi$ and $g_{\mu\nu}$ derived from the full action (including that of matter minimally coupled to $g_{\mu\nu}$), one can explicitly replace higher-order time derivatives of $\phi$ by at most second order time derivatives (see details in \cite{GLPV3}, and related ideas in \cite{Zumalacarregui:2013pma}). This shows that the equations of motion can be reduced to second order in time derivatives and do not require additional initial conditions, thus extending the conclusions of our Hamiltonian analysis to general configurations. The same method applies to theories without $L_4^\phi$, although one cannot simultaneously map  $L_4^\phi$ and $L_5^\phi$ to Horndeski for general combinations of these Lagrangians.

\vskip.1cm
\emph{Quadratic action.} 
The above arguments exclude the presence of extra DOF, but one still needs to check that the remaining scalar and tensor DOF are not themselves ghostlike, for which we need 
 to calculate the quadratic action for perturbations of the propagating DOF and make sure that the kinetic terms have the right signs. We perform this calculation around a spatially flat FLRW metric and follow the general procedure developed in \cite{GLPV}  for the specific Lagrangian $L$ given by eq.~\eqref{action}. Namely, we
 expand at second order the action $S=\int d^4 x\sqrt{-g} L$, using $\zeta$-gauge, \emph{i.e.}~$h_{ij} = a^2(t)e^{2 \zeta} (\delta_{ij} + \gamma_{ij} )$, $\gamma_{ii} = 0 = \partial_i \gamma_{ij}$, and splitting the shift as $N^i = \partial_i \psi + N_V^i$, $\partial_i N_V^i=0$. 
Because of the particular structure of the terms in eqs.~\eqref{K2} and \eqref{K3},  
the Lagrangian \eqref{action} satisfies the criteria obtained in \cite{GLPV} that ensure that the linear equations of motion contain no more than two spatial derivatives. In particular, terms proportional to 
$(\partial^2 \psi)^2$  cancel up to a total derivative.
By varying the action with respect to $N^i$, one obtains the momentum constraints, whose solution is $N_V^i =0$ and
\be
 N = 1+ {\cal D} \dot \zeta \;, \qquad  {\cal D} \equiv \frac{4{\cal A}_4}{ 2 H (2 {\cal A}_4+{\cal A}_4') -{\cal A}_3' } \;. \label{mc}
\ee
Above and in the following a dot and a prime, respectively, denote derivative with respect to $t$ and $N$. Furthermore, we use the new functions 
\be
\label{new}
\begin{split}
&{\cal A}_2\equiv A_2+3H A_3+6H^2 A_4+6H^3A_5\;,\\
&{\cal A}_3\equiv A_3+6H A_4 +12H^2 A_5\;,\\
&{\cal A}_4\equiv A_4+3H A_5\;,\\
&{\cal B}_4\equiv B_4+\frac{1}{2N} \dot B_5 |_{N=1}-(N-1) \frac{HB_{5}'}2 \;.
\end{split}
\ee
After substitution of eq.~\eqref{mc}  
 into the action 
all the terms containing $\psi$  drop out, up to boundary terms \cite{PV}. After some manipulations the quadratic action becomes $S^{(2)}=\int d^4 x \, a^3 {L}^{(2)}$ with
\be \label{lag-quad}
{L}^{(2)} = \alpha \dot \zeta^2  -  \beta \frac{(\partial_i \zeta)^2}{a^2} +  \frac{1}{4} \bigg[ - {\cal A}_4{\dot \gamma}_{ij}^2 - {\cal B}_4 \frac{(\partial_k \gamma_{ij})^2}{a^2}   \bigg]\;,
\ee
where the functions $\alpha$ and $\beta$ are defined as 
\be
\begin{split}
\alpha  \equiv & \left[ \frac{(N^2 {\cal A}_2')'}2 - 3H {\cal A}_{3}' + 6 H^2 (N{\cal A}_4)'  \right]    {\cal D}^2 -6 {\cal A}_4 \;, \\ 
\beta  \equiv &  - 2{\cal B}_4   + \frac{2}{a}\frac{d}{dt} \left[ a  {\cal D}   (N {\cal B}_4)'  \right] \;,
\end{split}
\ee
evaluated on the background ($N=1$). As expected from the previous Hamiltonian analysis, the quadratic Lagrangian~\eqref{lag-quad} does not contain higher time derivatives. Moreover, 
for $\alpha>0$ and $-{\cal A}_4>0$ we ensure that the  propagating DOF are not ghostlike. Gradient instabilities are avoided for $c_s^2 \equiv \beta/\alpha >0$ 
and $c_\gamma^2 \equiv -{\cal B}_4/{\cal A}_4>0 $.

\vskip.1cm
\emph{Coupling with matter.} In  cosmology, the power of gravity at large scales---and its irrelevance at short distances---is well illustrated by the Jeans phenomenon. 
A matter overdensity $\delta \rho_m$ of a given Fourier mode $k$ evolves, schematically, as 
\begin{equation} \label{schematic}
\left(\partial_t^2 \, +\,  c_m^2 k^2 \, -\,  {\rm gravity}\right)\  \delta \rho_m \ = \ 0\, .
\end{equation}
In the above, $c_m^2$ is the square of the speed of sound, proportional to the pressure perturbation, $c_m^2 = \delta p_m/\delta \rho_m$.  For $c_m^2 > 0$, the positive sign in front of the $k^2$ term guarantees an oscillating solution at sufficiently short distances, where the overdensity is supported by its own pressure gradients.  The last term in parentheses stands for $k$-independent contributions  roughly of Hubble size $\sim H^2$.  Only at  distances larger than $\sim c_m H^{-1}$ do these terms dominate, leading to gravitational (Jeans) instability. 
This well-known feature of standard cosmological perturbation theory holds true at small scales also in most modified gravity models---say, for definiteness, in all Horndeski theories as long as matter fields are minimally coupled to the metric.

The extension of Horndeski theories that we are proposing provides a  counterexample to such an apparently universal behavior, even when matter is minimally coupled to the metric tensor. Let us illustrate this with a matter scalar field $\sigma$ (not to be confused with the  dark energy field $\phi$), described by the $k$-essence type action,
\be
S_m=\int d^4x\sqrt{-g}\;P(\Y,\sigma)\,, \qquad  \Y \equiv g^{\mu \nu}\partial_\mu \sigma \partial_\nu \sigma\,,
\ee
with sound speed $c_m^2 \equiv {P_{,\Y}}/({P_{,\Y}-2\dot{\sigma}_0^2 P_{,\Y \Y}})$. 
One can then repeat the procedure discussed earlier in order to obtain the quadratic action for the scalar fluctuations expressed in terms of $\zeta$, $ N$, $\psi$ and the matter field perturbation $\delta\sigma$. Making use of the momentum constraints, the final Lagrangian expressed in terms of $\zeta$ and of the gauge-invariant variable
 $Q_\sigma \equiv \delta \sigma  - ({\dot \sigma_0}/{H})\zeta $,
reads
\begin{align}
& {L}^{(2)} =  \bigg( \alpha -  \frac{ c_m^2 g_t^2 }{4  P_{,\Y} } \bigg)  \dot \zeta^2  - \bigg( \beta  +  \frac{P_{,\Y} \dot \sigma_0^2}{H^2} - \frac{\dot \sigma_0 g_s }{H} \bigg) \frac{(\partial_i \zeta)^2}{a^2}  \nonumber \\
& - \frac{P_{,\Y}}{c_m^2}  \bigg( \dot Q_\s^2-c_m^2 \frac{(\partial_i Q_\s)^2}{a^2} \bigg) 
+ g_t \dot\zeta \dot Q_\s + g_s \frac{\partial_i \zeta\partial_i Q_\s}{a^2}  + \ldots
\;,
\end{align}
where  $g_s \equiv -c_m^2 g_t  +   2 \dot \sigma_0 P_{,\Y} \Delta $, with
\newcommand{\B}{{\cal B}}
\newcommand{\SK}{{\cal S}}
\newcommand{\sR}{{\cal R}}
\be
g_t �\equiv \frac{2 \dot{\sigma}_0P_{,\Y}}{c_m^2} \bigg( �{\cal D} -\frac{1 }{ H } \bigg), \quad �
\Delta � \equiv �{\cal D} \bigg( 1+ \frac{ � (N {\cal B}_4)' }{ {\cal A}_4 �} \bigg) \; 
\ee
and we have included only the terms quadratic in time or space derivatives, the other terms (in the ellipses)  being irrelevant for the following discussion.
The dispersion relations for the propagating DOF can be obtained by requiring that the determinant of the matrix of the kinetic and spatial gradient terms vanishes, which yields 
\begin{equation}
\begin{split}
&(\omega^2 - c_m^2 k^2) ( \omega^2 - \tilde c_s^2 k^2)  = \frac{ (\rho_m +p_m)}{ 2 \alpha} \, \Delta ^2  \, \omega^2 k^2 \, , \label{km} \\
&\tilde c_s^2 \equiv \big[\beta -  (1/2) (\rho_m +p_m)  ( {\cal D} -\Delta  )^2 \big]/{\alpha} \;,
\end{split}
\end{equation}
where we have used $2 \dot \sigma_0^2P_{,\Y} = -(\rho_m +p_m)$. From this equation one derives the two dispersion relations $\omega^2 = c^2_{\pm} k^2$.  In Horndeski theories, $\Delta \propto {\cal A}_4 + (N {\cal B}_4 )' =0$ because of eq.~\eqref{gal}, and 
we thus find that, despite the couplings in the action between the time and space derivative of $\zeta$ and $Q_\sigma$,  the matter sound speed is unchanged as a consequence of the special relation  $g_s = -c_m^2 g_t$. This is no longer the case in our non-Horndeski extensions,  where $\Delta\neq 0$ and the two couplings are ``detuned''. This remarkable difference between Horndeski and non-Horndeski theories was not pointed out in the recent work Ref.~\cite{Gergely:2014rna}, which also extends our previous analysis~\cite{GLPV} to compute the quadratic action of dark energy coupled to  a scalar field.

This unusual behavior can also be seen by writing the perturbed EOM  derived from the manifestly covariant Lagrangian for $\phi$, together with eq.~\eqref{schematic}. On sufficiently small scales, we  find (see \cite{GLPV3} for details)
\begin{align}
&(\partial_t^2  + \tilde c_s^2 k^2) \delta \phi - C_\phi  \dot \phi \, \partial_t \delta  \rho_m \approx 0 \;, \label{sfe} \\
&(\partial_t^2  + c_m^2 k^2) \delta \rho_m - C_m k^2 \, \partial_t (\delta  \phi/\dot \phi)  \approx 0  \;,
\end{align}
with 
\be
C_m  \equiv \frac{\Delta (\rho_m + p_m)}{ \Delta-{\cal D} }   \;, \qquad C_\phi  \equiv  - \frac{\Delta ( \Delta-{\cal D} )}{2 \alpha}  \;,
\ee
which leads to the same dispersion relation as in eq. (\ref{km}). This clearly shows  that, 
in contrast to the standard Jeans lore, the gravitational scalar  mode $\delta \phi$ cannot be decoupled from matter by going at sufficiently short distances. The origin of
the special coupling between  matter and the scalar field in eq.~\eqref{sfe} can also be understood as follows. Taking the example of $L_4$ for simplicity, one can see that the variation of  (\ref{L4}) with respect to $\phi$ yields a term of the form  $\phi^\lambda(g^{\mu \nu}+n^\mu n^\nu) \nabla_\nu R_{\lambda\mu}$.  Using Einstein's equations (this assumes to separate $L_4$  into a GR term and an effective additional term), one can express the Ricci tensor in terms of the matter energy-momentum tensor, which leads to the term $\dot\phi \, \partial_t \delta  \rho_m$ in eq.~\eqref{sfe}.

\vskip.1cm
\emph{Conclusion.} We have introduced a novel class of scalar-tensor theories, which include and extend Horndeski theories. For configurations where the  scalar field gradient is timelike, these theories can  be formulated in a very simple form via an ADM description of spacetime based on uniform $\phi$ slicing. 
This formulation allows to absorb the scalar degree of freedom in the spatial metric, and makes it particularly transparent to show the absence of  Ostrogradski instabilities.
For generic configurations, one can   use disformal transformations to relate subclasses of these theories to theories with manifest second-order equations of motion.  However,  this procedure cannot be simultaneously applied to the most general case that includes both $L_4^\phi$ and $L_5^\phi$, which means that a complete understanding of the full  covariant theory requires further investigation. 

An important corollary of this work applies to the original galileons proposed in~\cite{NRT}: Their direct covariantization, obtained by substituting ordinary derivatives with covariant ones, belongs to the class of theories considered here. Our work suggests that such theories are already free of ghosts instabilities and do not need the gravitational ``counterterms'' prescribed in~\cite{Deffayet:2009wt}.

We have also uncovered a remarkable phenomenological property of the non-Horndeski subclass of our theories:   
When {\em minimally} coupled to ordinary matter, they exhibit a kinetic-type coupling, leading   to a mixing of the dark energy and matter  sound speeds. It would be interesting to study further the phenomenology of these theories.

\vskip.1cm
\emph{Acknowledgements:} We  thank Dario Bettoni, Paolo Creminelli, Lorenzo Sorbo, Enrico Trincherini and especially Guido D'Amico and Andrew Tolley for enlightening discussions. 
F.P.~acknowledges the financial support of the UnivEarthS Labex program (ANR-10-LABX-0023 and ANR-11-IDEX-0005-02) and the A*MIDEX project (n. ANR-11-IDEX-0001-02) funded by the ``Investissements d'Avenir" French Government program, managed by the French National Research Agency (ANR). J.G.~and F.V.~acknowledge financial support from {\em Programme National de Cosmologie and Galaxies} (PNCG) of CNRS/INSU, France. D.L.~acknowledges financial support from the ANR (grant STR-COSMO ANR-09-BLAN-0157-01).

\appendix



\begin{thebibliography}{99}  

\bibitem{Clifton:2011jh} 
  T.~Clifton, P.~G.~Ferreira, A.~Padilla and C.~Skordis,
  ``Modified Gravity and Cosmology,''
  Phys.\ Rept.\  {\bf 513}, 1 (2012)
  [arXiv:1106.2476 [astro-ph.CO]].
    
    
  \bibitem{NRT}
A.~Nicolis, R.~Rattazzi and E.~Trincherini,
  ``The galileon as a local modification of gravity,''
  Phys.\ Rev.\ D {\bf 79}, 064036 (2009)
  [arXiv:0811.2197 [hep-th]].
  
     \bibitem{woodard} 
  R.~P.~Woodard,
  ``Avoiding dark energy with 1/r modifications of gravity,''
  Lect.\ Notes Phys.\  {\bf 720}, 403 (2007)
  [astro-ph/0601672].

 
 \bibitem{Deffayet:2009wt} 
  C.~Deffayet, G.~Esposito-Farese and A.~Vikman,
  ``Covariant galileon,''
  Phys.\ Rev.\ D {\bf 79}, 084003 (2009)
  [arXiv:0901.1314 [hep-th]].



\bibitem{Deffayet:2009mn} 
  C.~Deffayet, S.~Deser and G.~Esposito-Farese,
  ``Generalized galileons: All scalar models whose curved background extensions maintain second-order field equations and stress-tensors,''
  Phys.\ Rev.\ D {\bf 80}, 064015 (2009)
  [arXiv:0906.1967 [gr-qc]].
  
\bibitem{Deffayet:2011gz} 
  C.~Deffayet, X.~Gao, D.~A.~Steer and G.~Zahariade,
  ``From k-essence to generalised Galileons,''
  Phys.\ Rev.\ D {\bf 84}, 064039 (2011)
  [arXiv:1103.3260 [hep-th]].
  
    \bibitem{horndeski}
G.~W.~Horndeski, 
Int.\ J.\ Theor.\ Phys. {\bf 10}, 363 (1974).
 
 \cite{deRham:2011qq}
\bibitem{deRham:2011qq} 
  C.~de Rham, G.~Gabadadze and A.~J.~Tolley,
  ``Helicity Decomposition of Ghost-free Massive Gravity,''
  JHEP {\bf 1111}, 093 (2011)
  [arXiv:1108.4521 [hep-th]].
  
\bibitem{Zumalacarregui:2013pma} 
  M.~Zumalac\'arregui and J.~Garc\'ia-Bellido,
  ``Transforming gravity: from derivative couplings to matter to second-order scalar-tensor theories beyond the Horndeski Lagrangian,''
  Phys.\ Rev.\ D {\bf 89}, 064046 (2014)
  [arXiv:1308.4685 [gr-qc]].

\bibitem{Henneaux:1992ig} 
  M.~Henneaux and C.~Teitelboim,
  ``Quantization of gauge systems,''
  Princeton, USA: Univ. Pr. (1992) 520 p
  
  
\bibitem{GLPV3} 
J.~Gleyzes, D.~Langlois, F.~Piazza and F.~Vernizzi,
  ``Exploring gravitational theories beyond Horndeski,''
 JCAP {\bf 1502}, 02, 018 (2015)
  [arXiv:1408.1952 [astro-ph.CO]].

  




\bibitem{GLPV} 
 J.~Gleyzes, D.~Langlois, F.~Piazza and F.~Vernizzi,
``Essential Building Blocks of Dark Energy,''
JCAP {\bf 1308}, 025 (2013)
 [arXiv:1304.4840 [hep-th]].





\bibitem{PV} 
 F.~Piazza and F.~Vernizzi,
 ``Effective Field Theory of Cosmological Perturbations,''
 Class.\ Quant.\ Grav.\  {\bf 30}, 214007 (2013)
 [arXiv:1307.4350].


  \bibitem{Gergely:2014rna} 
  L.~\'A.~Gergely and S.~Tsujikawa,
  ``Effective field theory of modified gravity with two scalar fields: dark energy and dark matter,''
  Phys.\ Rev.\ D {\bf 89}, 064059 (2014)
  [arXiv:1402.0553 [hep-th]].
  
  
  
  
  


  

  
      

\end{thebibliography}
\end{document}